\documentclass[superscriptaddress,twocolumn,showpacs,prb,longbibliography,aps]{revtex4-1}
\usepackage[utf8]{inputenc}
\usepackage{graphicx}	
\usepackage{amsmath,bm,amssymb}	
\usepackage{color}

\usepackage{braket}
\providecommand{\abs}[1]{\lvert#1\rvert}
\providecommand{\norm}[1]{\lVert#1\rVert}

\makeatother

\begin{document}
\title{DFT+DMFT with natural atomic orbital projectors}

\author{Jae-Hoon Sim}
\affiliation{Department of Physics, Korea Advanced Institute of Science and Technology
	(KAIST), Daejeon 34141, Korea}

\author{Myung Joon Han}
\email{mj.han@kaist.ac.kr}
\affiliation{Department of Physics, Korea Advanced Institute of Science and Technology
	(KAIST), Daejeon 34141, Korea}

\date{\today}

\begin{abstract}
We introduce natural atomic orbitals as the local projector to define the correlated subspace for DFT + DMFT (density functional theory plus dynamical mean-field theory) calculation.
The natural atomic orbitals are found to be stably constructed against the number and the radius of basis orbitals.
It can also be self-consistently updated inside the  DFT+DMFT loop.
The spatial localization, electron occupation and the degree of correlation are investigated and compared with other conventional techniques. As a `natural' choice to describe the electron numbers, adopting natural atomic orbitals has advantage in terms of electron number counting. We further explore the reduction of computation cost by separating correlated orbitals into two subgroups based on the orbital occupancy. Our new recipe can serve as a useful choice for DFT+DMFT and related methods. 
\end{abstract}

\maketitle

\section{Introduction}

The calculation of real materials with strong electronic correlations poses an important problem in condensed matter physics and material science. While the first-principles band structure method based on density functional theory (DFT) provides a powerful theoretical framework for real materials, the sizable on-site correlation invalidates the standard approximations such as local density approximation (LDA) and generalized gradient approximation (GGA). To overcome this limitation, the combined methods of LDA/GGA with many-body techniques have been suggested and made tremendous success such as the early attempts of so-called LDA$+U$ or DFT$+U$ \cite{AnisimovAndersen_Band1991,AnisimovSawatzky_Densityfunctional1993} and the more recently suggested LDA+Gutzwiller \cite{HoWang_Gutzwiller2008,GutzwillerGutzwiller_Correlation1965,DengFang_Local2009,LanataKotliar_Phase2015}.
GW type of self energy within the self-consistent framework \cite{vanSchilfgaardeFaleev_Quasiparticle2006,KotaniFaleev_Quasiparticle2007,FaleevKotani_AllElectron2004} can also be useful with the limited purposes of, {\it e.g.}, describing the model parameters and the metallic Fermi surfaces \cite{JangHan_Quasiparticle2015,RyeeHan_Quasiparticle2016}.
Among them, DFT+DMFT (dynamical mean-field theory) has become one of the standard choices providing the unique feature by capturing the `dynamic' correlations
\cite{GeorgesKotliar_Hubbard1992, MetznerVollhardt_Correlated1989,ZhangKotliar_Mott1993,
	AnisimovKotliar_Firstprinciples1997, LichtensteinKatsnelson_Initio1998,
	KotliarMarianetti_Electronic2006,GeorgesRozenberg_Dynamical1996}.

Along with its great success,
		many of formal and technical issues in the implementation  have received much attention \cite{AmadonLichtenstein_Planewave2008, AnisimovVollhardt_Full2005, SavrasovAbrahams_Correlated2001,HeldScalettar_Cerium2001,HeldAnisimov_MottHubbard2001, ParkKotliar_Cluster2008, deMediciGeorges_JanusFaced2011, KentKotliar_Predictive2018,ChoiKotliar_ComDMFT2019,_DMFTpack,AletWessel_ALPS2005,ParcolletSeth_TRIQS2015, HauleKim_Dynamical2010}.
Typically, any attempt to combine LDA/GGA-type of scheme with
Hubbard-like model approach requires a step to define the correlated subspace within which Coulomb interaction (`Hubbard $U$') comes in to play. In other words, the correlated orbitals span the bands near the Fermi level ($E_F$) where the on-site correlation becomes important. They are expected to be reasonably well localized, site-centered and atomic-like. In literature, many different suggestions to define the correlated orbital space are found:  maximally localized Wannier functions (MLWF) \cite{SouzaVanderbilt_Maximallylocalized2001,MarzariVanderbilt_Maximally1997}, muffin-tin orbitals \cite{AndersenSaha-Dasgupta_Muffintin2000,PavariniAndersen_Mott2004},
and other projection methods \cite{HauleKim_Dynamical2010,HauleKotliar_Covalency2014}.

Considering the ambiguity in this choice of `projector', we take a special note of recent studies that emphasize the correlated orbital occupancy $N_{d}$ being
the critical variable to describe the correlation effect \cite{WangMillis_Covalency2012,DangMarianetti_Covalency2014}.
Here it should first be noted that estimating or determining $N_{d}$ is a  non-trivial issue. It depends on the form of so-called double counting form \cite{WangMillis_Covalency2012,DangMarianetti_Covalency2014}
as well as the choice of the correlated orbitals \cite{HauleKotliar_Covalency2014}.
In fact, it is not straightforward already in DFT-LDA level, leading to many different possible choices suggested for atomic orbitals or charge counting schemes such as Mulliken population analysis  \cite{MullikenMulliken_Electronic1955,MullikenMulliken_Electronic1955a}, Löwdin orthogonalization method \cite{LowdinLowdin_Nonorthogonality1950}
and natural population analysis \cite{ReedWeinhold_Natural1985}.

In this paper we suggest the implementation of DFT+DMFT with natural atomic orbitals (NAOs) as the local projector. 
In our DFT formulation based on the linear combination of non-orthogonal pseudo-atomic orbitals (PAOs) \cite{openmx}, the use of NAOs provides
a `natural' way of estimating $N_{d}$ from the orthogonalized orbitals.
The NAO construction process is not only well plugged in between DFT and DMFT self-consistent loop, but it also
stable with respect to the numerical parameters such as
the range of energy window and the number of basis sets.
Our calculations show that the localization and the covalency are reasonably well described with NAO compared with the other choices including PAO, L${\rm \ddot{o}}$wdin orthogonalized orbital (LOO), and MLWF.
Further, the possibility of reducing computation cost is explored by separating the correlated subspace into two different parts.

The paper is organized as follow. In Sec.~\ref{sec:Formalism} we
briefly review the DFT+DMFT procedure which is followed by our NAO-based formalism.
In Sec.~\ref{sec:Application}, the calculation results are presented for
a correlated metal SrVO$_{3}$ and a charge-transfer insulator
NiO. Summary and conclusion are given in Sec.~\ref{sec:Conclusion}.

\section{Formalism\label{sec:Formalism}}

\subsection{DFT+DMFT and the basis issue}

\begin{figure*}
\includegraphics[width=1.5\columnwidth]{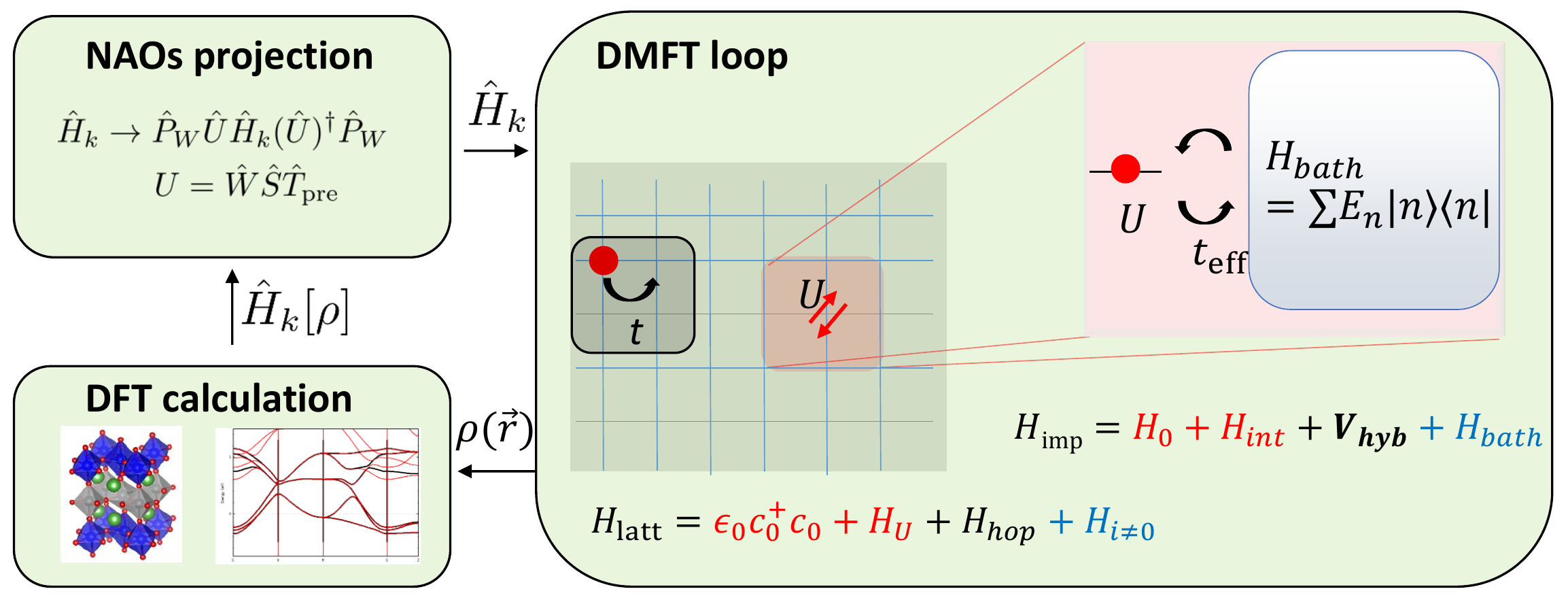}
\caption{A schematic illustration of our  DFT+DMFT calculation by using NAO as local projector. Kohn-Sham DFT Hamiltonian is solved within LDA or GGA which provides the charge density and overlap matrix. With those, one can construct NAOs which then serve as correlated orbitals for DMFT impurity model. As in the conventional DFT+DMFT, Kohn-Sham Hamiltonian $H_k$ obtained from LDA/GGA is regarded as non-interacting $H_0$ for DMFT calculation. The computed Green’s function $G_{{\rm imp}}$
and the self-energy $\Sigma_{{\rm imp}}$ complete the self-consistent loop by providing the updated charge density $\rho(\vec{r})$ for DFT-LDA/GGA.\label{fig:Schematic-diagram-for}}
\end{figure*}

So-called DFT+DMFT is a scheme that combines the standard
band theory such as LDA and GGA with a non-perturbative many-body technique, DMFT.
In DMFT, the lattice self-energy
$\hat{\Sigma}_{{\rm latt}}(i\omega_{n}, \boldsymbol{k})$ is
approximated by local self-energy 
$\hat{\Sigma}^{\rm DMFT}_{{\rm latt}}(i\omega_{n})=\bigoplus_R \ \hat{\Sigma}_{{\rm loc}} (R)$.
The non-local contribution is projected out
by a local projection; $\hat{\Sigma}_{\rm loc} (R) = P_{R}\hat{\Sigma}_{{\rm latt}}=\sum_{\alpha\beta\in d}\ket{\chi_{\alpha R}}\Sigma_{\alpha\beta}(R)\bra{\chi_{\beta R}}$
where `$d$' refers to the correlated subspace such as transition-metal $d$ orbitals and $R$ the atomic position.
The on-site Coulomb interaction is applied onto the localized orbitals $\ket{\chi_{\alpha}}$. The interacting
Green's function is given by
\begin{equation}
\hat{G}(i\omega_{n}, \boldsymbol{k})=\frac{{1}}{i\omega_{n}+\mu-\hat{H}(\boldsymbol{k)}-\hat{\Sigma}_{{\rm loc}}(i\omega_{n})+\hat{\Sigma}_{dc}(i\omega_{n})}
\end{equation}
where $\hat{\Sigma}_{dc}$ refers to the double-counting term.
Once $P_{R}$, or correlated subspace $\{\ket{\chi_{\alpha}}\}$
is specified, $\hat{\Sigma}_{{\rm loc}}$ is determined by solving
the impurity model with the self-consistently-constructed hybridization
function \cite{ZhangKotliar_Mott1993,LichtensteinKatsnelson_Initio1998,GeorgesKotliar_Hubbard1992,MetznerVollhardt_Correlated1989}:
\begin{widetext}
\begin{equation}
\hat{\Delta}(i\omega_{n})=P_{R=0}\left[i\omega_{n}-\hat{H}_{{\rm loc}}-\Sigma_{{\rm loc}}(i\omega_{n})+\Sigma_{{\rm dc}}(i\omega_{n})-\left(\frac{1}{N_{k}}\sum_{k}\hat{G}(\boldsymbol{k},i\omega_{n})\right)^{-1}\right].
\end{equation}
\end{widetext}

Many issues can arise in solving this problem. The first thing is to choose `impurity solver' for which several standard techniques are available such as 
(continuous-time) quantum Monte Carlo ((CT)QMC) \cite{WernerMillis_ContinuousTime2006,WernerMillis_Hybridization2006},
exact diagonalization \cite{KochGunnarsson_Sum2008,CaffarelKrauth_Exact1994},
NRG (numerical renormalization group) \cite{WilsonWilson_Renormalization1975,BullaPruschke_Numerical2008}
and others \cite{BauernfeindEvertz_Fork2017}, each of which has both advantage and disadvantage. Another important issue is related to $\hat{\Sigma}_{dc}(i\omega_{n})$
for which many different recipes have been discussed in literature
\cite{WangMillis_Covalency2012,KarolakLichtenstein_Double2010,HauleHaule_Exact2015,HauleHaule_Exact2015,AnisimovAndersen_Band1991,AnisimovSawatzky_Densityfunctional1993}
In the current study, we take so-called `fully localized limit' as suggested in Ref.~\onlinecite{AnisimovSawatzky_Densityfunctional1993}.

Our main concern here is about the projection for defining the correlated
subspace $\left\{ \ket{\chi_{\alpha}}\right\} $ in the line of previous discussion \cite{WangMillis_Covalency2012,HauleKotliar_Covalency2014,DangMarianetti_Covalency2014}. Note that different projection method can lead to the different self-energy.
Several different ways to construct localized orbitals have been suggested.
For example, maximally localized Wannier
function (MLWF) method is widely used in combination with different type of codes \cite{LechermannAndersen_Dynamical2006,BhandaryHeld_Charge2016,_DCore}.
One can also take 
LMTO (linearized muffin-tin orbitals) \cite{PourovskiiGeorges_Selfconsistency2007}, NMTO (n-th order muffin-tin orbitals) \cite{LechermannAndersen_Dynamical2006},
or resort to the real space embedding method \cite{HauleKim_Dynamical2010}.

As an alternative possible choice, we pay a special attention to `projective Wannier function (PWF)' method which is a straightforward way to define correlated orbitals \cite{AmadonAmadon_Selfconsistent2012,AmadonLichtenstein_Planewave2008}. It has been adopted in some standard computation schemes including LMTO \cite{AnisimovVollhardt_Full2005}, APW \cite{AichhornBiermann_Dynamical2009,KarolakLichtenstein_General2011,SchulerAichhorn_Charge2018a}, and PAW \cite{AmadonAmadon_Selfconsistent2012,AmadonLichtenstein_Planewave2008}. The main idea of PWF is to project the localized atomic-like orbitals $\left\{ \ket{\tilde{\chi}_{\alpha}}\right\}$
onto the the low-energy Bloch states 
\[
\ket{\chi_{\alpha}}=\frac{{1}}{N}\sum_{k,n}^{\epsilon_{kn}\in W}\ket{kn}\braket{kn}{\tilde{\chi}_{\alpha}},
\]
 followed by orthogonalization procedure \cite{AmadonLichtenstein_Planewave2008}.
Here $N=\norm{\sum_{k,n}^{\epsilon_{kn}\in W}\ket{kn}\braket{kn}{\tilde{\chi}_{\alpha}}}$
is the normalization factor and $W$ energy window within which the hybridization function $\Delta(i\omega_n)$ is defined. Thus the result depends not only on the energy window $W$ but also on the choice of initial
local orbitals $\left\{ \ket{\tilde{\chi}_{\alpha}}\right\} $.
While the ambiguity related to the $W$ range can be removed, at least in principle by taking the large enough energy window \cite{HauleKotliar_Covalency2014}, the choice of 
the $\ket{\tilde{\chi}_{\alpha}}$ is not straightforward. The
different initial choice can lead to the different final result
\cite{KarolakLichtenstein_General2011}.

A purpose of our current work is to construct the optimal $\ket{\tilde{\chi}_{\alpha}}$ for the non-orthogonal local orbital basis method.
 As mentioned above, the result of DFT+DMFT can critically depend on the choice of $\ket{\tilde{\chi}_{\alpha}}$.
Non-orthogonality can introduce an ambiguity in the estimation of key physical quantities such as
the number of electrons in $d$ orbitals $N_{d}$. This poses a highly non-trivial issue not only because the correlated orbitals often get significantly hybridized with ligands but also because the numerical orthogonalization process usually introduces
the undesirable mixing between the two. Here we suggest NAOs \cite{ReedWeinhold_Natural1985} as a set of the local correlated orbitals $\ket{\tilde{\chi}_{\alpha}}$ for DFT+DMFT calculation. As illustrated in Fig.~\ref{fig:Schematic-diagram-for}, this process can be inserted as an intermediate step between DFT and DMFT to complete the self-consistent loop. While `natural orbital' has been adopted for the basis set to solve the impurity model \cite{KananenkaZgid_Systematically2015}, we emphasize that the use of $\ket{\chi_{\alpha}^{{\rm NAOs}}}$ as $\ket{\tilde{\chi}_{\alpha}}$ has never been reported for the first-principles DFT+DMFT.

The use of NAO as a local projector has notable advantages. As will be discussed in the below, this choice can certainly give 
the more intuitive charge counting for the correlated orbitals. Further, the construction of NAOs is found to be numerically stable. In the case of MLWF, for example, achieving the convergence can be an issue when the bands are strongly entangled. As for muffin-tin orbitals or PAOs, basically the similar  ambiguity can be manifested in terms of the choice of local orbital radius. In this sense, the use of NAO can be a `natural' choice regardless of the DFT basis types.

\subsection{Natural atomic orbital as a projector}

The numerical PAOs basis in our DFT code, OpenMX, is constructed in a controlled way and therefore its localization is well identified \cite{OzakiKino_Numerical2004,OzakiOzaki_Variationally2003}.
The non-orthogonality issue, on the other hand, needs to be dealt with care. One straightforward way is just to take $\ket{\tilde{\chi}_{\alpha}}=\ket{\phi^{{\rm PAOs}}}$ as correlated orbitals and to reconstruct the ligand orbitals to be orthogonal with respect to $\ket{\tilde{\chi}_{\alpha}}$ through, e.g., Schmidt orthogonalization procedure.
This choice corresponds to \textit{`full'} local projector in the
previous DFT$+U$ implementation \cite{HanYu_LDA2006}. 
Not surprisingly, however, this procedure overestimates the electron occupation in the correlated orbitals as discussed in the below (see Sec. \ref{sec:Application}).
Another possible choice, namely, Löwdin symmetric orthogonalization, is not the desirable method either because it treats the important atomic set (the AO with large occupations)
and the less important AO (the almost empty AO) on the equal footing.

Here we note that NAOs can be determined in a physically motivated way and suitable for local orbitals of the given materials \cite{ReedWeinhold_Natural1985}. 
Mathematically, NAOs are defined by the eigen-orbitals of any given local occupation matrix \cite{LowdinLowdin_Quantum1955}.
For a given occupation matrix $N^{\rm PAOs}_{\alpha\beta}=\braket{c_{\alpha}^{\dagger}c_{\beta}}$ NAOs are constructed through a three-step process.
 In the below, orbital index $\alpha  \equiv (ilp)$ specifies the site index $i$,
	angular momentum quantum number $l$, and multiplicity number of radial basis function $p$. 
First, we construct the atom-centered local orbitals $\ket{\phi^{{\rm pre}}}$ called as `pre-NAOs' which are
the eigenstates of the sub-block $(N^{i,l})_{p p^\prime}$
corresponding to the occupation matrix projected onto the atomic site $i$ and the angular momentum $l$ subshells. To preserve the invariance of the occupation with respect to
the coordinate transformation, the symmetry averaging should be carried out over the $(2l+1)$ diagonal blocks of $N_{p p^\prime}^{i,l}$. Since the pre-NAO transformation matrix $\hat{T}_{\rm pre}^{i,l}$
considers only the sub-blocks of the occupation matrix, pre-NAOs centered on different atoms are nonorthogonal to one another.

The second and third step concern about the proper elimination of inter-atomic wave function overlaps. To obtain a stable result, we divide pre-NAOs into two subsets, namely, `minimal'
and `Rydberg' set \cite{ReedWeinhold_Natural1985}. The `minimal' set is the atomic $(n,l)$ subshells with finite formal occupancy whereas the `Rydberg' set consists of the remaining
(formally unoccupied) orbitals. Then the `Rydberg' sets are Schmidt
orthogonalized to the manifold spanned by `minimal' orbitals. We represent this orthogonalization by a matrix $\hat{S}$. This step is essential to avoid the over-counting problem in the occupancy-weighted symmetric orthogonalization (OWSO) process in the next step.

In the third step, we orthogonalize pre-NAOs orbitals by means of OWSO method in which the occupancy-weighted difference between orthogonalized
basis and original pre-NAOs
\begin{equation}
\sum_{\alpha}N_{\alpha\alpha}\norm{\ket{\chi_{\alpha}^{{\rm NAOs}}}-\ket{\phi_{\alpha}^{{\rm pre}}}}^{2}\label{eq:OWSO}
\end{equation}
is minimized. Here, $N$ is expressed  within pre-NAOs basis. The transformation having this property is written as 
\begin{equation}
\ket{\chi_{\alpha}^{{\rm NAOs}}}=\ket{\phi_{\beta}^{PAOs}}W_{\beta\alpha},\label{eq:OWSO_Trans}
\end{equation}
where $\hat{W}=\tilde{N}(\tilde{N} O \tilde{N})^{-1/2}$ with the overlap matrix of pre-NAOs ($O$), and the diagonal part of $N$ ($\tilde{N}$). 
The final form of transformation matrix from
PAOs to NAOs can then be written as  $\hat{T}_{N}=\hat{W}\hat{S}\hat{T}_{Pre}$.

Natural orbital methods, including NAO and natural bond orbital, have been used typically to calculate atomic charge population \cite{ReedWeinhold_Natural1985, ReedWeinhold_Intermolecular1988}.
Recently, natural orbitals have also been used to solve impurity problem of strongly correlated electron systems \cite{GoMillis_Adaptively2017,MaydaBulut_Electronic2017}. For example, natural orbitals can provide the adaptive basis set to reduce the dimension of Hilbert space \cite{GoMillis_Adaptively2017}. In Ref.~\onlinecite{MaydaBulut_Electronic2017}, model parameters for Anderson impurity problem have been obtained based on NAOs. In the current study, we used NAO as a basis or a projector for representing correlated subspace. We also demonstrate that the use of NAO can provide a way of reducing computation costs by separating the correlated orbitals into two subsets and adopting two different levels of solvers as will be discussed in the below.

\section{Application\label{sec:Application}}

\subsection{Computation details}

First-principles band calculations have been carried
out based on DFT within LDA \cite{PerdewZunger_Selfinteraction1981,CeperleyAlder_Ground1980}.
We used `OpenMX’ code \cite{openmx,OzakiKino_Numerical2004,OzakiOzaki_Variationally2003}
for DFT calculation. Experimental lattice parameters have been used
\cite{ReyCyrot-Lackmann_Preparation1990, FurstenauLangell_Initial1985}, and the $k$-point meshes of $13\times13\times13$ and $23\times23\times23$ adopted for SrVO\textsubscript{3} and NiO, respectively. Double valence and single polarization orbitals were used as a basis set for DFT. These numerical atomic basis orbitals ({\it i.e.}, PAOs) were generated with cutoff radius of 10.0, 6.0, 6.0, and 5.0 a.u. for Sr V, Ni, and O atoms, respectively \cite{OzakiKino_Numerical2004}.
The DFT-LDA calculation results serve as the non-interacting Hamiltonian $H_{0}$, and the interaction Hamiltonian
containing the density-density interaction is parameterized by $U$ and $J_{H}$
for on-site Coulomb repulsion and Hund interaction,
respectively. The Hamiltonian is solved within the single-site DMFT
by employing hybridization expansion CT-QMC algorithm
\cite{GullWerner_Continuoustime2001,GullTroyer_Continuoustime2011}.
For `impurity solver', we used the software package as implemented in Ref.~\onlinecite{HauleHaule_Quantum2007, _CTQMC}, and the results were double checked by using ALPS library
\cite{AletWessel_ALPS2005}.
Our DFT+DMFT interface code is available online
\cite{_DMFTpack}.
The real frequency self-energy and spectral function are obtained from the matsubara Green’s function and the self-energy by analytic continuation
using maximum entropy method (MEM) \cite{JarrellGubernatis_Bayesian1996,GunnarssonSangiovanni_Analytical2010a}.
For this process we used our recent method as reported in Ref.~\onlinecite{SimHan_Maximum2018}.
The large enough energy windows of $W=[-10, 10]$ eV
around $E_F$ has been considered \cite{ReyCyrot-Lackmann_Preparation1990,HauleKotliar_Covalency2014}.
	For comparison, we also present the results of MLWF as local projectors where
   the initial projections onto three atomic V-$t_{2g}$ and nine O-$2p$ orbitals were used for SrVO$_3$  \cite{SouzaVanderbilt_Maximallylocalized2001, MarzariVanderbilt_Maximally1997}.

\subsection{SrVO\protect\textsubscript{3} \label{subsec:SVO113} }

\begin{figure}
\includegraphics[width=1\columnwidth]{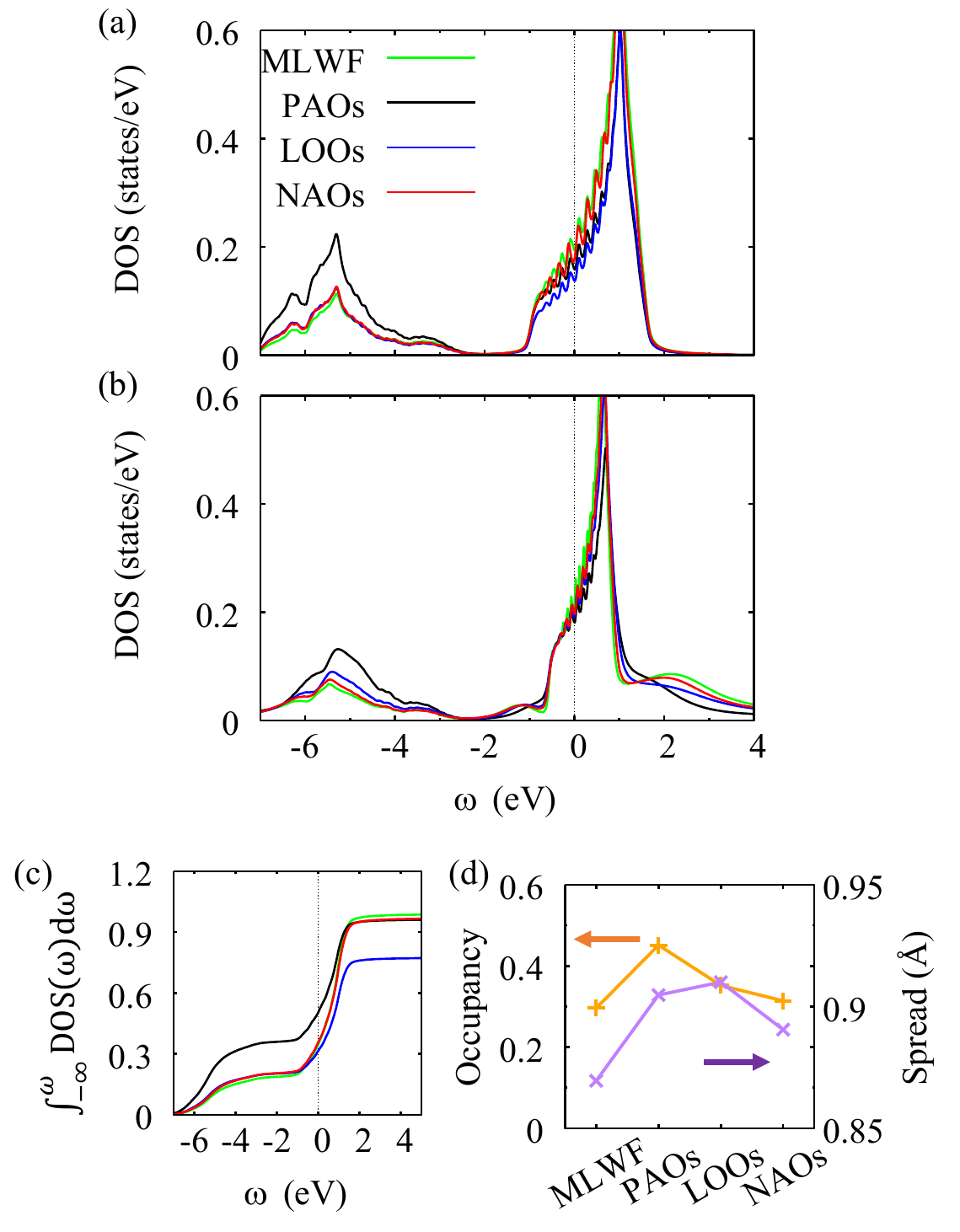}
\caption{
	(a) The calculated density of states projected onto the V-$t_{2g}$-like orbitals for SrVO$_{3}$ with $U=0$ eV. 
	The results from the different local orbitals, namely MLWF (green), PAOs (black), LOOs (blue), and NAOs (red), are compared. 
	(b) The calculated spectral function $A(\omega)$ projected onto the local orbitals with $U$ = 6 eV and
	at inverse temperature $\beta=20$ eV\protect\textsuperscript{-1}.
	(c) The integrated density of states for different local orbitals.
	(d) The calculated occupancy of each $t_{2g}$ orbitals within LDA+DMFT calculation (orange) and the spread of the orbitals as defined in the text (purple).
\label{fig:SVO_U0_PDOS}}
\end{figure}

SrVO$_{3}$ has been serving as
a test bed for DFT+DMFT \cite{NekrasovAnisimov_Comparative2005,NekrasovVollhardt_Momentumresolved2006,KarolakLichtenstein_General2011,SekiyamaAnisimov_Mutual2004,NomuraImada_Effective2012} and the related methods such as 
LDA+DCA \cite{LeeValenti_Dynamical2012} and $GW$+DMFT \cite{TarantoHeld_Comparing2013,CasulaBiermann_Dynamical2012,TomczakBiermann_Asymmetry2014,TomczakBiermann_Combined2012,ChoiKotliar_Firstprinciples2016}.
With cubic SrVO$_{3}$ as our first example, we investigated the properties of NAO as a local projector. 
Fig.~\ref{fig:SVO_U0_PDOS}(a) and (b) shows the LDA ($U=0$ eV) and the LDA+DMFT density of states (DOS) projected onto the V-$t_{2g}$ like orbitals, respectively. Two main peaks are clearly identified; the O$_p$--V$_d$ bonding complex locating at around $[-7 {\rm eV}, -3 {\rm eV}]$, and the the antibonding states across $E_F$ which are mainly of V$_d$ character. The results of four different projectors are presented in different colors, namely, the green, black, blue and red lines refer to the MLWF, PAO, LOO, and NAO result, respectively.

Not surprisingly, the degree of hybridization depends on the choice of projectors.
For example, the direct use of PAOs notably overestimates $d$-$p$ hybridization compared to the other projection methods. This feature is reflected in that the more states of the lower-lying bonding complex are assigned as V$_d$ orbitals; see Fig.~\ref{fig:SVO_U0_PDOS}(a) and (b). Also, it naturally affects the electron number counting. With PAO projector,
$n_{e_g}^{\rm LDA}=\sum_{n,k}^{\epsilon_{nk}<\epsilon_{F}}
\abs{\braket{nk}{\tilde{\chi}_{\alpha}^{{\rm PAO}}}}^{2}=0.55$
and 
$n_{t_{2g}}^{\rm LDA}=\sum_{n,k}^{\epsilon_{nk}<\epsilon_{F}}
\abs{\braket{nk}{\tilde{\chi}_{\beta}^{{\rm PAO}}}}^{2} =0.50$, where $\alpha\in e_g$ and $\beta \in t_{2g}$.
Both values of $n_{t_{2g}}^{\rm LDA}$ and $n_{e_g}^{\rm LDA}$ are notably larger than
	those of NAOs; $n_{e_g}^{\rm LDA}= 0.29$ and $n_{t_{2g}}^{\rm LDA}=0.35$.
The nominal values ($t_{2g}^1$ configuration of V$^{4+}$) are $n_{e_g}=0$ and $n_{t_{2g}}=1/6$. Note that $n_{t_{2g}}^{\rm LDA} < n_{e_g}^{\rm LDA}$ in PAO projection whereas
$n_{t_{2g}}^{\rm LDA} > n_{e_g}^{\rm LDA}$ in NAO.

The same feature can also be noted in the integrated DOS (IDOS) 
as presented in Fig \ref{fig:SVO_U0_PDOS}(c). The calculated IDOS at $E_F$ 
clearly shows that the V$_d$ occupation or its valency is markedly larger in the PAO estimation than in others. The calculated IDOS at $E=E_F$ is 0.34, 0.50, 0.31, and 0.35 for MLWF, PAO, LOO, and NAO, respectively. The largest value of PAO projector reflects the greatest $d$--$p$ hybridization while the value of NAO is smaller than that of PAO and comparable with MLWF. Another notable feature is that the IDOS calculated from LOOs does not reach to but remains far below 1.0 even up to $\omega=+10$ eV; $\simeq$ 0.78. On the other hand, the use of PAO and NAO projector satisfies the sum rule, yielding the
IDOS $\simeq$0.96 and 0.97 ({\it i.e}., close to 1.0), respectively, as $\omega$ becomes large.

This feature remains stable even when the number of basis  orbitals (not the projector orbitals) changes.
	The calculation with the more basis orbitals of $s2p2d2f$$1$ ({\it i.e.}, double orbital sets for $s$, $p$, $d$, and single for $f$ states), for example, gives $n_{t_{2g}}^{\rm LDA}=0.31$ by using NAO projector.
The calculated IDOS up to high energy remains as 0.97. On the other hand, the IDOS result of LOO, 0.72, shows the sizable dependence on the basis set choice due to the mixing between the correlated orbitals near $E_F$ and the atomic orbitals in the higher energy. These features can certainly be useful for both practically and physically.

In order to see and compare the degree of spatial localization of correlated orbitals produced by different projectors, we computed the `spread function' \cite{MarzariVanderbilt_Maximally1997} defined by
$\Omega  = \sqrt{{\braket{\mathbf{r}^{2}}-\braket{\mathbf{r}}^{2}}}$
where $\braket{\mathbf{r}} = \int  \mathbf{r} \abs{\tilde{\chi}_ {\alpha\in t_{2g}}(\mathbf{r})}^2 d\mathbf{r}$ $~$.
The results of the energy window $W=[-10,10]$ eV are presented in Fig.~\ref{fig:SVO_U0_PDOS}(d).
The calculated $\Omega$ for NAOs is 0.891~{\AA} which is slightly larger than that of MLWF (0.870~{\AA}), and smaller than PAO ($0.905$~{\AA}) and LOO ($0.910$~{\AA}). Our analysis shows that the NAO projector produces a moderately localized spatial subspace of correlation.

The DFT+DMFT result of $A(\omega)=-1/\pi\Im G_{{\rm loc}}(\omega)$ is presented in Fig.~\ref{fig:SVO_U0_PDOS}(b). The calculations were performed with the inverse temperature $\beta=20$ eV$^{-1}$, $U=6.5$ eV and $J_{H}=0.65$ eV \cite{AmadonLichtenstein_Planewave2008}.
The effect of correlation is clearly seen in the bandwidth renormalization and the upper Hubbard-like peak developed at around $\omega=+2$eV, both of which clearly show that the correlation effect is gradually reduced as the more localized projector is adopted. The use of NAO gives the moderate degree of correlation in between MLWF and PAO. Focusing on the lower and upper Hubbard peak, identified at around $\omega=-1$ and $+2.5$ eV, respectively, they are more pronounced in the result of MLWF than PAO. Once again, the result of NAO is in the middle being consistent with the above analysis.
A systematic trend of correlation strength depending on the local projectors is also observed in the calculated spectral weight of the bonding orbital complex.The calculated $A(\omega)$ with PAO projector has the much greater weight in this region of energy  reflecting the larger hybridization between V-$t_{2g}$ and O-$p$ states. This is attributed to the extended and non-orthogonal nature of PAO.

Fig.~\ref{fig:SVO_U0_PDOS}(d) also shows the electron occupancy in V$_{t_{2g}}$ orbitals, $n_{t2g}^{\rm DMFT}$, obtained from LDA+DMFT calculation with different local projectors. The NAO shows the comparable value with that of MLWF which is noticeably smaller than the other projection results. 
The use of more localized projector results in the smaller occupation which is consistent with the previous studies on covalency issue 
\cite{WangMillis_Covalency2012,DangMarianetti_Covalency2014,HauleKotliar_Covalency2014}.

A straightforward and quantitative way to measure the correlation effect is to estimate the quasi-particle renormalization factor 
\begin{equation}
Z\approx\left[1-\frac{\Im\Sigma(i\omega_{1})}{\omega_{1}}\right]^{-1}.
\end{equation}
The calculated value by NAO projector is $Z^{\rm NAO}=0.60$ which is comparable with that of MLWF ($Z^{\rm MLWF}=0.58$). The result of PAO and LOO is 
$Z^{{\rm PAOs}}=0.62$ and $Z^{{\rm L\ddot{o}wdin}}=0.61$, respectively. While these calculation results are in overall good agreement with the previous DFT+DMFT calculation of 0.61   \cite{KarolakLichtenstein_General2011} and
ARPES (angle-resolved photoemission spectroscopy) of 0.56 \cite{YoshidaShen_Direct2005}, the degree of correlation effect is once again follow the same trend discussed above.

\begin{figure}
	\includegraphics[width=0.9\linewidth]{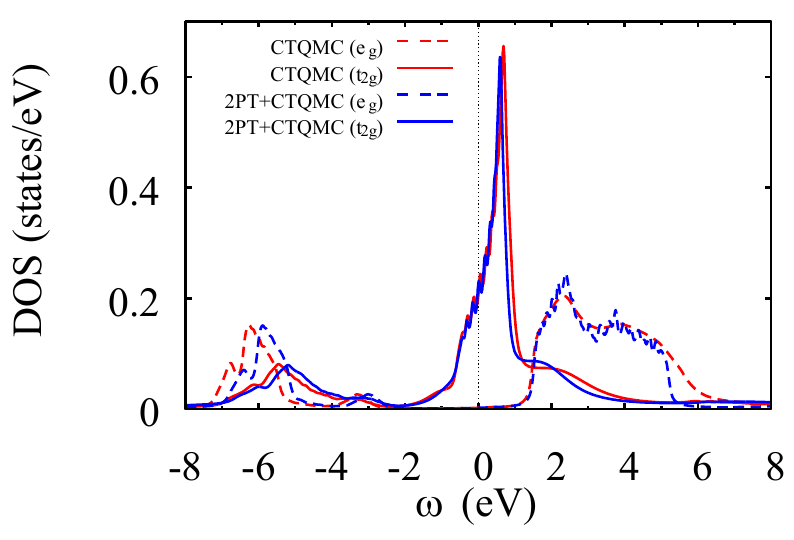}
	\caption{
		The calculated spectral function for SrVO$_3$ projected onto the V-$e_{\rm g}$ (dashed line) and $t_{\rm 2g}$ (solid line) orbitals. The result presented in red color is obtained with
		CTQMC solver for five V-$d$ orbitals. The blue colored lines present the result obtained by CTQMC solver for three V-$t_{\rm 2g}$ orbitals whose self-energy is then embedded into the second-order perturbation theory (2PT) solution. \label{fig:SVO_5orbital_dos}
	}
\end{figure}

One big advantage of using our NAO projector is to reliably separate the correlated subspace into two parts. It enables us to use an elaborate technique only for the one part of correlated orbitals while, for another part, a computationally cheaper approximation can be utilized. Or, high-level approximation can be adopted for
both parts, which are then embedded by the self-energy obtained from the cheaper approximation as shown in recent model studies  \cite{KananenkaZgid_Systematically2015, NguyenLanZgid_Rigorous2016, LanZgid_Testing2017}.
In this scheme, natural orbitals were used as a basis set to represent the correlated subspace by taking only the orbitals whose occupations are close to 1.0.

We apply this type of capability to the first-principles DFT+DMFT framework. In order to see its effect, we extend the correlated subspace from three $t_{2g}$ orbitals to the $d$ complex ({\it i.e.}, both $t_{2g}+e_g$). The former is certainly more relevant to the correlation effect and basically determines the most of electronic properties while the latter is less important. Therefore we adopted CTQMC solver for $t_{2g}$ orbitals and the second-order perturbation theory (2PT) for $e_g$, which significantly reduces the computation cost. For more details of our 2PT method, see Appendix~\ref{app-A}.

The calculation result of spectral function is presented as a blue line in Fig.~\ref{fig:SVO_5orbital_dos}. The red line in Fig.~\ref{fig:SVO_5orbital_dos} represents the CTQMC-only result; namely, all of five V-$d$ orbitals are solved with CTQMC. In spite of much less computation cost ({\it i.e.}, five- vs three-orbital impurity problem for CTQMC), the hybrid solver of 2PT+CTQMC gives a reasonable agreement with the CTQMC-only result especially for the near-$E_F$ region. For example, Hubbard bands and the $d$-$p$ hybridization part are well reproduced. Simultaneously, some deviations are also clearly noticed. For example, the intensity of the Hubbard bands are reduced in 2PT+CTQMC calculation. This reduced correlation in $t_{\rm 2g}$ manifolds is attributed to the inability of 2PT to accurately describe the screening effect \cite{GukelbergerWerner_Dangers2015, HonerkampWerner_Limitations2018a}. As expected, the difference between the two computation results is more pronounced in high energy $e_g$ spectra.

\subsection{NiO}

\begin{figure}
\includegraphics[width=0.8\linewidth]{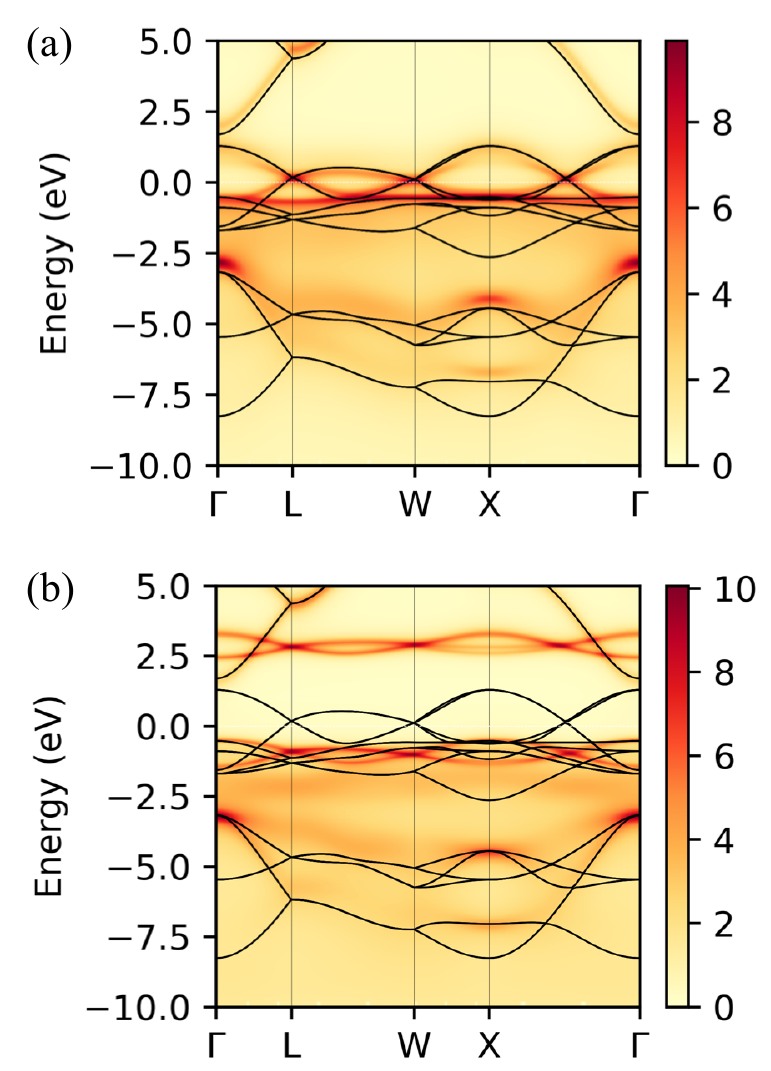}
\caption{The calculated band dispersion of  NiO. DFT+DMFT result (color map) is compared with that of LDA (black solid lines). 
	Two different projection
methods are adopted for DFT+DMFT calculation; (a) PAOs and (b)
NAOs.\label{fig:Band_structure_NiO} }
\end{figure}

As the second example, we chose a classical charge-transfer insulator  NiO. Due to the sizable hybridization between Ni-$d$ and O-$p$, the electronic property depends on the choice of local correlated orbital projector. Note that the projector affects both Hubbard $U$ and charge-transfer energy $\Delta$ in this type of materials \cite{ZaanenAllen_Band1985,WangMillis_Covalency2012,DangMarianetti_Covalency2014}. The inverse temperature of $\beta=10$ eV$^{-1}$ (corresponding to 1160.45 K higher than Neel temperature $T_{N}=525$ K \cite{SrinivasanSeehra_Nature1983,RothRoth_Multispin1958}) and
the interaction parameters of
$U=8$ eV and $J_H=1$ eV were used following the
previous studies \cite{KarolakLichtenstein_Double2010,KunesVollhardt_Local2007,AnisimovAndersen_Band1991}. Fig.~\ref{fig:Band_structure_NiO} shows the electronic structures which were  calculated by two different projectors;
(a) PAO and (b) NAO. As expected, LDA ($U$=0; black solid line) gives the unphysical metallic solution while the experimental gap is $\sim$4.3 eV \cite{HufnerHulliger_Photoemission1984}.

The DFT+DMFT spectral function $A({\bf k},\omega)$ calculated with
PAO projector is presented as a false-color band in Fig.~\ref{fig:Band_structure_NiO}(a). Interestingly the system remains metallic in a sharp contrast to the previous calculations of DFT+$U$ \cite{AnisimovAndersen_Band1991,HanYu_LDA2006}
and DFT+DMFT \cite{KunesVollhardt_Local2007,KarolakLichtenstein_Double2010}.
This result demonstrates the fact that the effect of correlations depend on the choice of local projector. Note that the Ni-${e_g}$ energy level
$\epsilon^{e_g}=\braket{\hat{H}}_{\chi^{\rm PAOs}}=-7.58$ eV 
is significantly lower than that of NAO (see below) which
results in the large number of Ni-$d$ electrons, 
$N_{d}=8.72$. In combination with the conventional FLL double
counting, it leads to a significant change of chemical potential.
This effect is clearly noticed, for example, in the bands at around $\pm\sim$2.5 eV being deviated from LDA bands.

The result of NAO projector is presented in 
Fig.~\ref{fig:Band_structure_NiO}(b) (false color plot) in comparison with LDA (black line). A well developed band gap is clearly identified. The on-site energy is significantly increased compared to PAO result, $\epsilon_{d}^{eg}=-7.21$ eV, and therefore the charge transfer
energy is increased. The $d$ orbital occupancy is also reduced $N_{d}=8.21$.
It is found that the electronic structure is overall quite consistent with previous
theoretical studies \cite{HanYu_LDA2006,KarolakLichtenstein_Double2010,KunesVollhardt_Local2007}.

\section{Summary\label{sec:Conclusion}}

We introduce NAO as a local projector to define the correlated subspace in DFT+DMFT procedure. Our implementation is based on the systematic construction of projector from the original non-orthogonal PAO basis set. We apply this method to a correlated metal SrVO\textsubscript{3} and insulator NiO. From the comparison with other projector methods,
we found that NAO not just serves as another possible choice, but it also has some advantage particularly in the charge counting. First, it provides a reliable electron number $N_d$ for the correlated orbitals and does not require any additional convergence or minimization procedure. No arbitrary numerical parameter needs to be introduced such as cut-off or muffin-tin radius, and the application to the entangled band structure is also straightforward. Finally, we emphasize that the use of NAO projector provides a viable way to separate correlated subspace into two parts; one for which the elaborate technique can be used for describing correlations, and for another part one can resort to a computationally cheaper approximation.

\begin{acknowledgments}
	We thank Hongkee Yoon for useful comment and discussion.
	This work was supported by the Basic Science Research Program through the National Research Foundation of Korea (NRF) funded by the Ministry of Education(2018R1A2B2005204) and Creative Materials Discovery Program through the NRF funded by Ministry of Science and ICT (2018M3D1A1059001).
\end{acknowledgments}

\appendix

\section{Details of the second-order perturbation approach}
\label{app-A}

Our DFT+DMFT iteration procedure with 2PT+CTQMC solver is summarized as follow.

(1) We start with the initial guess for self-energy $\Delta\Sigma(i\omega_{n})=\Sigma_{\text{weak}}+\Sigma_{\text{strong}}-\Sigma_{\text{DC}}^{\text{DFT}}=0$. The local Green's function of a given material (or lattice problem) is
\begin{equation}
G_{\text{loc }}(i\omega_{n})=\frac{1}{N_{k}}\sum_{k}\left[i\omega_{n}-H_{k}+\mu-\Delta\Sigma(i\omega_{n})\right]^{-1}
\end{equation}
where $H_{k}$ is the Kohn-Sham Hamiltonian and the chemical potential
$\mu$ is adjusted to obtain the correct number of electrons.

(2) Calculate Weiss mean-field $\mathcal{G}^{-1}=G_{\text{loc}}^{-1}+\Sigma_{\text{imp}}$.
Equivalently one can calculate the impurity energy level and the hybridization
function from $H_{\text{imp}}=H_{\text{loc}}-\Sigma_{\text{DC}}^{\text{DFT}}$
and $\Delta=i\omega_{n}-H_{\text{loc}}+\mu-\Delta\Sigma-G_{\text{loc}}^{-1}$, respectively.

(3) Now we solve the impurity problem with an approximate way. Here we adopted
the second-order perturbation theory:
\begin{equation}
\Sigma^{(2)}(i\omega_{n})=\Sigma^{\text{HF}}[G_{\text{loc}}]+\Sigma^{\prime(2)}[G_{\text{HF}}](i\omega_{n}),
\end{equation}
where $\Sigma^{\text{HF}}$ is the Hartree-Fock contribution and
 	\begin{widetext}
\begin{equation}
\Sigma_{ij}^{\prime(2)}[G](i\omega_{n})= (-G_{kl}(\tau) G_{mn}(\tau) G_{pq}(-\tau) + G_{kn}(\tau) G_{ml}(\tau) G_{pq}(-\tau)  )  U_{iqmk}U_{lnpj}.
\end{equation}
\end{widetext}
Here the summation over the repeated indices is assumed and $U_{ijkl} = \braket{ij | U | kl}$.

(4) Calculate the double counting term, $\Sigma_{\text{DC}}^{(2)}=\Sigma^{\text{HF}}[\hat{P}_{c}G_{\text{loc}}]+\Sigma^{\prime(2)}[\hat{P}_{c}G_{\text{HF}}](i\omega_{n}).$
Here $\hat{P}_{c}$ is the projector onto the strongly correlated
orbitals, namely, $t_{\text{2g}}$ orbitals for SVO\textsubscript{3}. Then we have $\Sigma_{\text{weak}}(i\omega_{n})=\Sigma^{(2)}(i\omega_{n})-\Sigma_{\text{DC}}^{(2)}(i\omega_{n})$.

(5) CTQMC impurity solver is adopted to obtain the impurity self-energy
$\Sigma_{\text{strong}}$ describing the correlated subspace $c$.
Again, we construct the impurity problem from the Weiss field $\mathcal{G_{\text{strong}}}^{-1}=(P_{c}G_{\text{loc}})^{-1}+\Sigma_{\text{strong}}$.
Alternatively, we can define the impurity site energy $H_{\text{imp}}^{\text{strong}}=H_{\text{loc}}+\Sigma_{\text{weak}}(\infty)-\Sigma_{\text{DC}}^{\text{DFT}}$
and the hybridization $\Delta_{\text{strong}}=i\omega_{n}-H_{\text{imp}}^{\text{strong}}+\mu-\Sigma_{\text{strong}}-(P_{c}G_{\text{loc}})^{-1}\approx i\omega_{n}-H_{\text{loc}}+\mu-P_{c}(\Delta\Sigma)+\Sigma_{\text{weak}}^{\prime}-(P_{c}G_{\text{loc}})^{-1}.$

(6) Update $\Delta\Sigma(i\omega_{n})=\Sigma_{\text{weak}}+\Sigma_{\text{strong}}-\Sigma_{\text{DC}}^{\text{DFT}}$
and go back to step 2.


%

\end{document}